**Ultra-dense phosphorus in germanium delta-doped layers.**

G. Scappucci,[1]

*School of Physics and Australian Research Council Centre of Excellence for Quantum Computer Technology, University of New South Wales, Sydney, NSW 2052, Australia.*

G. Capellini,

*Dipartimento di Fisica, Università di Roma Tre, Via della Vasca Navale 84, 00146 Roma, Italy.*

W. C. T. Lee, and M. Y. Simmons

*School of Physics University of New South Wales, Sydney, NSW 2052, Australia.*

**Abstract**

Phosphorus (P) in germanium (Ge) $\delta$-doped layers are fabricated in ultra-high vacuum by adsorption of phosphine molecules onto an atomically flat clean Ge(001) surface followed by thermal incorporation of P into the lattice and epitaxial Ge overgrowth by molecular beam epitaxy. Structural and electrical characterizations show that P atoms are confined, with minimal diffusion, into an ultra-narrow 2-nm-wide layer with an electrically-active sheet carrier concentration of $4\times10^{13}$ cm$^{-2}$ at 4.2 K. These results open up the possibility of ultra-narrow source/drain regions with unprecedented carrier densities for Ge n-channel field effect transistors.

[1] Electronic mail: giordano.scappucci@unsw.edu.au; michelle.simmons@unsw.edu.au

On the roadmap of continuous device miniaturization beyond the 22 nm node, quasi-ballistic operation with injection at the source-end is necessary to maintain high speed operation in transistors.[1] To accomplish this, technological progress other than silicon (Si) scaling is necessary, including the use of alternative high mobility channel materials. Germanium (Ge) is a promising candidate for the replacement of Si in the next generation of nanoscale metal oxide semiconductor field effect transistors (MOSFETs) because the higher bulk mobility (~ 3× higher for electrons and ~ 5× for holes at room temperature) and the higher saturation velocity (~ 7×) lead to maximum drain current while minimizing leakage currents and power dissipation.[2]

Although the earliest transistor and integrated circuits were both fabricated in Ge, further progress in Ge MOSFETs was hindered by the lack of a stable native oxide that could passivate the surface and act as a high quality gate dielectric. Prompted by the renewed interest in Ge-based nanoelectronics, the advancement on high permittivity (high-$\kappa$) dielectrics has now enabled the fabrication of high-quality p-channel Ge MOSFETs.[3] The development of satisfactory n-channel Ge MOSFETs, however, is facing a major roadblock. Due to the enhanced diffusion of n-type dopants in Ge compared with bulk Si, forming reliable n-type source-drain contacts within the thermal budget allowed by high-κ dielectrics ($T < 400$ °C) has become a critical issue.[4] Traditional ion implantation has been so far inadequate to meet these requirements due to high temperatures required for electrical activation of dopants. In this letter we present an ultra-high vacuum (UHV), low thermal budget ($T \leq 350$ °C) technique enabling the fabrication of P $\delta$-layers in Ge and we demonstrate that $\delta$-doping is, also in a Ge matrix, a viable route for achieving high electrically-active n-type dopant concentrations with ultra-narrow profiles.

The $\delta$-layers fabrication strategy comprises an *ex-situ* chemical cleaning and *in-situ* surface preparation, adsorption of phosphine ($PH_3$) molecules on the clean Ge surface, incorporation of the P atoms into the Ge lattice and encapsulation of the $\delta$-layer with epitaxial Ge grown by low temperature (LT) molecular beam epitaxy (MBE). A similar approach was demonstrated for P $\delta$-doping of Si and has been integrated in an STM-based device scheme for the realization of Si devices with atomic scale dopant profiles.[5] In this paper we exploit the imaging capability of the scanning tunneling microscope (STM) at each stage of the $\delta$-doping process and then we characterized the $\delta$-layers *ex-situ* by secondary ion mass spectrometry (SIMS), transmission electron microscopy (TEM) and magnetotransport measurements.

The Ge $\delta$-doped layers were fabricated in an UHV STM-MBE system with a base pressure < $5 \times 10^{-11}$ mbar equipped with a Ge evaporator and a $PH_3$ dosing system. Samples with a size of $2.5 \times 10$ mm$^2$ were cleaved from a Sb doped Ge(001) 4 inch wafer (resistivity of 1-10 $\Omega$ cm). A clean atomically flat Ge(001) surface was prepared using a two-step method of *ex-situ* chemical passivation followed by an *in-situ* heating procedure. A $GeO_2$ passivation layer was chemically grown *ex-situ* by a wet treatment consisting of a $HCl:H_2O$ 36:100 bath and subsequent $H_2O_2:H_2O$ 7:100 bath to strip/reform a $GeO_2$ passivation layer. The samples were then outgassed *in-situ* at 230 °C for ~ 1 h, flash-annealed at 760 °C for 60 s to remove the $GeO_2$ and slowly cooled ( at ~ 2 °C s$^{-1}$) from 600 °C to room temperature to obtain an ordered reconstructed surface. Figure 1(a)-(f) shows a step by step schematic with corresponding STM images of the process for fabricating Ge:P $\delta$-doped layers in UHV. The as-prepared Ge(001) surface (Figure 1(b)) is atomically flat, with the typical (2×1) and higher order c(2×2) and p(2×4) reconstructions visible.[6] The sample was then heated to 100 °C to enhance the reactivity of the surface and dosed with $PH_3$ at a pressure of $5 \times 10^{-9}$ mbar for 10 min.. P atoms were incorporated into the surface by slowly heating the sample from 100 °C to 350 °C in 5 min. followed by a rapid cool-down to room temperature. The surface (Figure 1(d)) now shows interruption of the well ordered Ge(001) reconstruction and more contrast due the increased height difference between neighboring atoms, a signature of charge transfer from the incorporated P atoms, that form complex bonding configurations on Ge(001) as shown in previous high-resolution core-level photoemission experiments.[7] A detailed STM study of the interaction of $PH_3$ with the Ge(001) surface and of the P incorporation process goes beyond the scope of this letter and will be published elsewhere. The sheet of P atoms was then encapsulated with ~ 25 nm of intrinsic Ge deposited at a rate of 0.13 Å/s by MBE. The deposition temperature was kept as low as 210 °C to limit dopant redistribution at the atomic scale. STM investigation on a larger area of the resulting surface (Figure 1(f)) reveals the formation of small rounded mounds, with a peak to valley roughness of ~ 2 nm, as observed in previous studies of LT Ge homoepitaxy.[8] A close-up of the same surface (inset of Figure 1(f)) shows dimer rows running perpendicular to each other on alternate layers highlighting the epitaxial nature of the growth.

The structural properties of the $\delta$-layer were investigated *ex-situ* by SIMS and TEM. The $^{31}$P depth profile determined by SIMS (Figure 2(a)) was carried out with a Cs$^-$ primary ion beam at a low energy of 1 keV to optimize depth resolution. We attribute the sharp and isolated $^{31}$P peak 22 nm below the surface to the Ge:P $\delta$-layer. The concentration at peak maximum is $1\times10^{21}$ cm$^{-3}$ and has a full width at half maximum ~ 2 nm, smaller than the average ~ 5 nm Bohr radius for P donors in Ge, demonstrating the effectiveness of the doping technique in confining a two-dimensional (2D) sheet of P atoms in the starting Ge surface. The peak has a leading edge (1.2 nm/decade) steeper than the trailing edge (2.6 nm/decade) implying that very little diffusion occurred during the LT encapsulation process. The rise and decay rate are unavoidable well-known measurement artifact resulting from ion beam mixing from the sputtering process which make the peak appear broader than it really is. Additionally, the total concentration of P atoms incorporated in the $\delta$-layer was measured by integrating the depth profile yielding a value of $2.2\times10^{14}$ cm$^{-2}$ which corresponds to 0.35 monolayer (ML) density of P atoms incorporated into the Ge surface.

Figure 2(b) shows a TEM micrograph of the $\delta$-layer. Two arrows mark the expected position of the substrate/epilayer interface, 22 nm below the surface as suggested from the position of the $^{31}$P concentration peak. There is no evidence of interface between substrate and epilayer. This highlights the cleanliness of the starting surface and excludes the presence of strain due to highly dense P-incorporation. The $\delta$-layer doesn't interrupt the crystalline structure and the overgrowth is epitaxial throughout the whole encapsulation layer with no detected amorphous phase, indicating high-quality homoepitaxy. The contrast observed close to the surface is due to roughness - in agreement with the STM image reported in Figure 1(f) - and to defects, stacking faults and consequent cusps, that originate during the LT growth.[8] These, however, do not disturb the P atoms sheet buried ~ 20 nm under the surface where the Ge matrix has a high crystal quality, as observed in the TEM images (see Fig. 2(b)).

The electrical properties of the $\delta$-layers were studied by magnetotransport measurements at 4.2 K with standard low-frequency lock-in techniques. Trench-isolated Hall bars were defined by a CHF$_3$/CF$_4$ based dry etch and ohmic contacts to the $\delta$-layer were formed by thermally evaporated Al.[9] All measured Hall bars show ohmic conductivity with a sheet resistance of $\rho_{xx}$ = 5.5 k$\Omega$/□. The Hall resistance $\rho_{xy}$ has a linear dependence vs. perpendicular magnetic field $B$ (Figure 3(a)) and from the slope we calculate an electrically-active sheet carrier concentration of $n_{2D}$ = 4×10$^{13}$ cm$^{-2}$. Comparing this to the total incorporated P density 2.2 × 10$^{14}$ cm$^{-2}$ obtained from SIMS, we conclude that only 20% of the P atoms in the $\delta$-layer are electrically-active. Similar low activations were reported previously for Si:P $\delta$-layers heavily doped above the solid-solubility limit (SSL) for P in Si, and were ascribed to the formation of inactive P-related precipitates or dopant vacancy complexes which are known to reduce electrical activation.[10] The same might happen in the Ge:P layers since the SIMS profile shows a peak concentration of incorporated P of 1×10$^{21}$ cm$^{-3}$, well above the usually quoted SSL of 2×10$^{20}$ cm$^{-3}$ for P in Ge.[11] We expect to obtain higher activation by optimizing the dose/incorporation step in order reduce the amount of incorporated P to 0.25 ML needed for complete substitutional doping.

Despite the low percentage of activation, we achieve a remarkable density of electrically-active P dopants. We estimate the electrically-active 3D bulk concentration $N$ = 2×10$^{20}$ cm$^{-3}$ from $N \sim n_{2D}/w$, where $w$ is the 2 nm full width at half maximum of the SIMS $^{31}$P peak.[12] This value approaches the solubility limit and is ~ 4 times higher than other electrically-active concentration reported previously for P-doped Ge layers (~ 5×10$^{19}$ cm$^{-3}$).[4] From $n_{2D}$ and the zero field sheet resistance $\rho_{xx}(0)$ we calculate the electron mobility $\mu$ = 28.3 cm$^2$/Vs, elastic scattering time $\tau$ = 3.5×10$^{-15}$ s and mean free path $l$ = 1.5 nm. Low $\mu$ and $l$ are expected in ultra-dense $\delta$-layers due to the impurity scattering of the electrons strongly confined in the dopant layer.[10] Future attempts to enhance *µ* and *l* will include *in-situ* and/or *ex-situ* anneal steps to improve the crystal quality after encapsulation, as demonstrated for Si:P ultra-dense *δ*-layers.[13] Figure 3(b) shows a plot of the magnetoconductivity $\Delta\sigma(B) = [\rho_{xx}(B)]^{-1} - [\rho_{xx}(0)]^{-1}$, where $\rho_{xx}(B)$ is the measured longitudinal resistivity as a function of *B*. The inverted peak at zero field is due to the weak localization (WL) that arises from the coherent backscattering of electrons in time-reversed trajectories. Observation of WL is a clear signature of the 2D nature of transport in these systems and is associated with the strong confinement of the carriers in the *δ*-doped layer. For a disordered 2D system of non interacting electrons the WL correction to the conductivity *σ* is described by the Hikami formula:[14]

$$\delta\sigma_{WL}(B) = -\frac{\alpha e^2}{\pi h}\left[\psi\left(\frac{1}{2} + \frac{\hbar}{2ev_F^2\tau^2 B}\right) - \psi\left(\frac{1}{2} + \frac{\hbar}{2ev_F^2\tau\tau_\phi B}\right)\right] \quad (1)$$

where *α* is a phenomenological prefactor expected to be close to 1, *Ψ* is the digamma function, $\tau_\phi$ the phase relaxation time and $v_F$ the Fermi velocity. Together with the experimental data, in Figure 3(b) we report the theoretical fit of *Δσ*(B) to *δσ*$_{WL}$(B)−*δσ*$_{WL}$(0) using *α* and $\tau_\phi$ as fitting parameters. We found a best fit for *α* = 0.87 and $\tau_\phi$ = 5.6×10$^{-12}$ s and calculate a phase coherence length $l_\phi$ = 41.5 nm from $l_\phi^2 = \frac{1}{2}v_F^2\tau\tau_\phi$. This value is comparable to phase coherence values reported in Si:P *δ*-layers with comparable carrier densities.[10]

In summary, we have demonstrated that P doping at a high concentration is possible in Ge using a UHV compatible low thermal budget *δ*-doping technique. These results open up the possibility of forming reliable n-type source/drain regions in ultra-scaled Ge n-MOSFET. Following our recent demonstration of UHV-STM nanolithography on hydrogen-terminated Ge,[15] we intend to integrate this *δ*-doping technique in an STM-based device scheme to fabricate devices in Ge with atomic scale dopant profiles.

GS acknowledges support from UNSW under the 2009 Early Career Research and Science Faculty Research Grant scheme. GC is thankful to UNSW for a Visiting Professor Fellowship. MYS acknowledges an Australian Government Federation Fellowship.

**Figures**

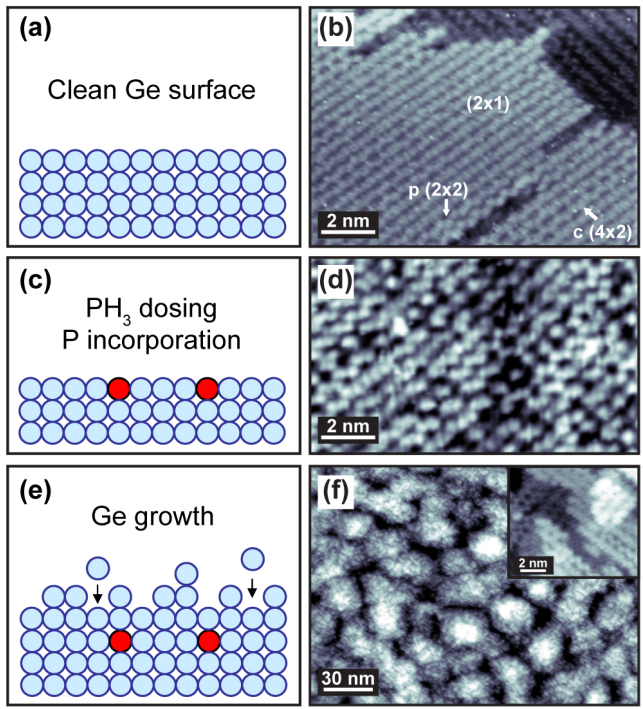

Figure 1:

(Color online) Schematic diagrams and corresponding STM images of the UHV fabrication process for a Ge:P $\delta$-doped layer. See text for details.

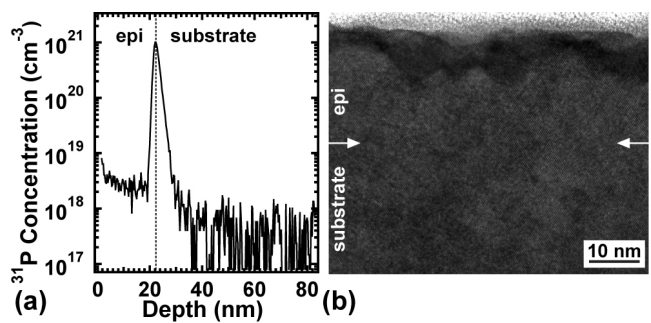

Figure 2:

(a) $^{31}$P depth profile of the Ge:P $\delta$-doped layer determined by SIMS and (b) TEM micrograph of the same $\delta$-layer, with arrows marking the expected interface between substrate and epilayer.

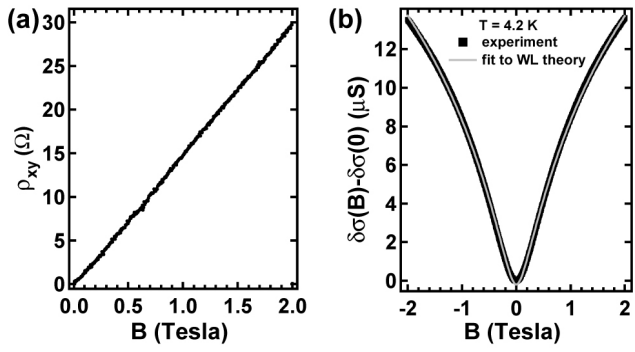

Figure 3:

(a) Hall resistance and (b) magnetoconductance (experimental and weak localization theoretical fit) of the Ge:P δ-doped layer as a function of perpendicular magnetic field measured at $T$ = 4.2 K.